\documentclass[sigconf,screen]{acmart}
\usepackage[shortlabels]{enumitem}
\AtBeginDocument{%
  \providecommand\BibTeX{{%
    \normalfont B\kern-0.5em{\scshape i\kern-0.25em b}\kern-0.8em\TeX}}}

\setcopyright{none}
\renewcommand\footnotetextcopyrightpermission[1]{} 

\begin{document}
\pagestyle{plain} 
\title{A Brief Survey of Open Radio Access Network (O-RAN) Security}

\author{YI-ZIH CHEN}
\email{r10946002@ntu.edu.tw}
\affiliation{%
  \institution{National Taiwan University}
  \city{Taipei}
  \country{R.O.C}}
\author{PO-JUNG SU}
\email{b08902078@ntu.edu.tw}
\affiliation{%
  \institution{National Taiwan University}
  \city{Taipei}
  \country{R.O.C}}
\author{TERRANCE YU-HAO CHEN}
\email{yhterrance@csie.ntu.edu.tw}
\affiliation{%
  \institution{National Taiwan University}
    \city{Taipei}
  \country{R.O.C}}
\author{CHI-TING LIU}
\email{b07901090@ntu.edu.tw}
\affiliation{%
  \institution{National Taiwan University}
    \city{Taipei}
  \country{R.O.C}}


\begin{abstract}
  Open Radio Access Network (O-RAN), a novel architecture that separates the traditional radio access network (RAN) into multiple disaggregated components, leads a revolution in the telecommunication ecosystems. Compared to the traditional RAN, the proposed O-RAN paradigm is more flexible and more cost-effective for the operators, vendors, and the public. The key design considerations of O-RAN include virtualization and intelligent capabilities in order to meet the new requirements of 5G. However, because of the open nature and the newly imported techniques in O-RAN architecture, the assessment of the security in O-RAN architecture during its early development stage is crucial. This project aims to present an investigation of the current ORAN architecture from several attack surfaces, including (1) Architectural openness, (2) Cloud and Virtualization, (3) Network slicing, and (4) Machine Learning. The existing attack surfaces and corresponding mitigation methods of these attacks are also surveyed and provided in this report, serving as a guiding principle and valuable recommendation for the O-RAN implementers and framework designers.
\end{abstract}


\begin{teaserfigure}
  \includegraphics[width=\textwidth]{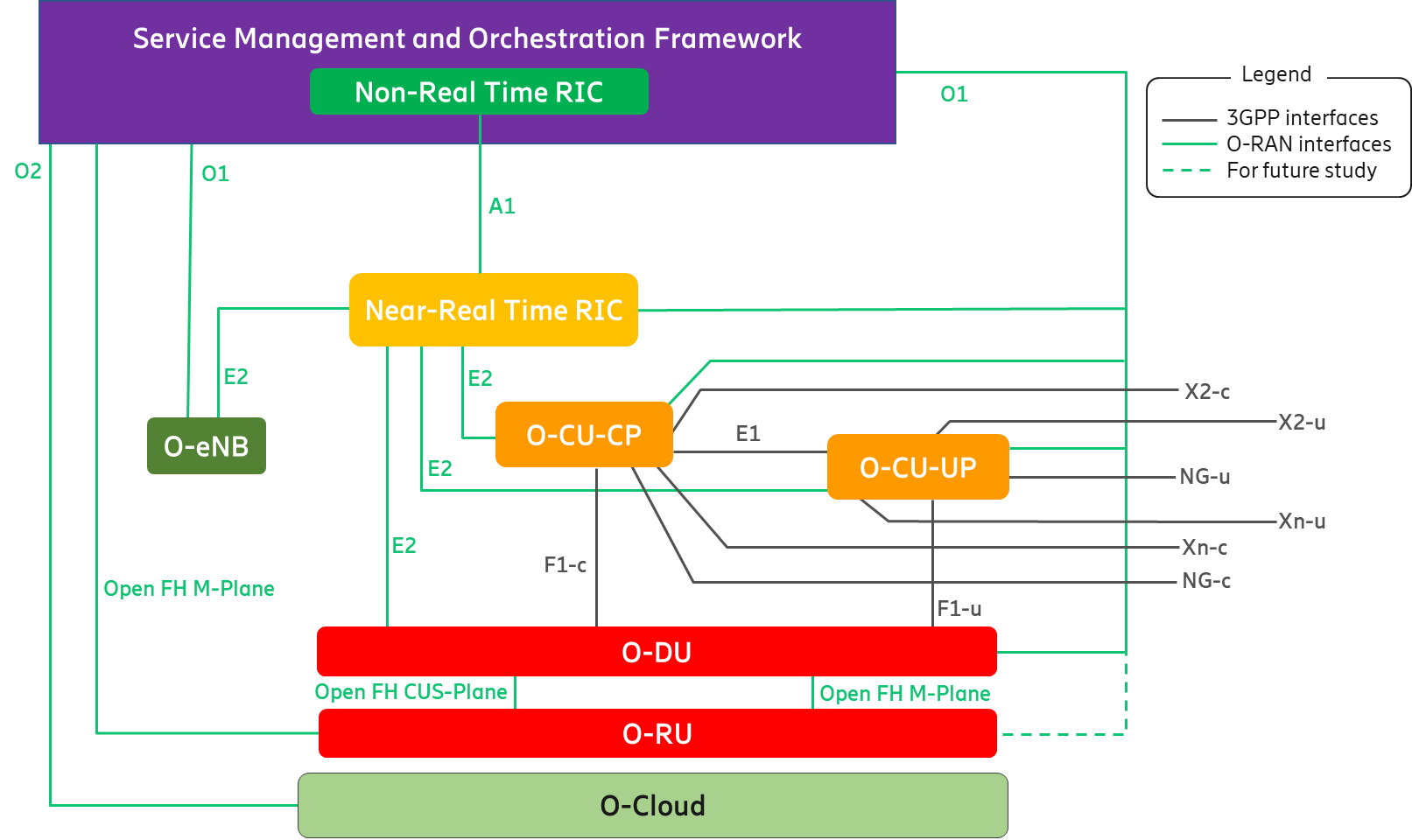}
  \caption{Logical Architecture of O-RAN proposed by O-RAN alliance}
  \label{fig:teaser}
\end{teaserfigure}

\maketitle

\section{Introduction} 

Next-generation telecommunication technologies aim to satisfy the
new demands for enhanced performance, portability, elasticity
and energy efficiency of heterogeneous services.  5G and 6G networks adopt architectural transformations to build a flexible, agile, and disaggregated system, which makes for the evolution of wireless networks. Among all the new concepts, the Open Radio Access Network (O-RAN) framework has been one of the most important innovations in the field to serve as the prime infrastructure for future mobile networks.

The objective of ORAN is to build a flexible, reliable, and cost-effective
RAN system by decomposing the Radio Access Network (RAN) into several
open and capable units based on cloud-native technology \cite{DBLP:journals/corr/abs-2201-06080, klement2022open}. Each component in ORAN is connected through open and standardized interfaces, making it interoperable across different vendors and third-party applications. Moreover, the ORAN framework also considers the integration of machine learning technologies for better network resource management such as resource allocation for consumers with different needs for quality of service (QoS). 

Fig. \ref{fig:teaser} shows the O-RAN framework proposed by the O-RAN alliance. Near the bottom of the schematic, we could see the O-cloud component, which serves as a cloud-based computing platform and hosts the functions such as the RAN Intelligent Controller (RIC), Centralized Unit (CU), and Distributed Unit (DU).
Located in the middle of the schematic are the CU and DU, which are derived from the conventional 3GPP-specified NG-RAN (Next Generation NodesBs Radio Access Network) that is responsible for the digital data processing computation. The splitting of CU and DU is based on the scenario of usage and the trade-offs between performance and bandwidth. CU could be further divided into the control plane (CP) and the user plane (UP) functions, which are also referred to as the CU-CP and CU-UP. This division, known as the Control User Plane Separation (CUPS), is based on various performance needs and could improve the RAN function placements.
Service Management and Orchestration (SMO) Framework and the RAN Intelligent Controllers (RICs) are shown at the top of the schematic. These functions aim to realize the automatic resource management and control of the O-RAN network. The Non-Real-Time RIC (Non-RT RIC) resides within SMO, supporting intelligent RAN management with policies and ML-based algorithms. On the other hand, the Near Real-Time RIC (Near-RT RIC) enables the near-real-time decision-making of O-RAN networks with a logical function or by leveraging the ML model trained on Non-RT RIC. The time response of the O-DU control loop is less than 10 milliseconds. For the Near-RT RIC and Non-RT RIC, the time response is between 10 milliseconds to 1 second and more than 1 second respectively.

Dissecting the RAN architecture results in the extensions of additional interfaces in the network, which could be identified as green lines in Fig. 1. 
The O1-interface is responsible for the network management tasks (FCAPS: Fault, Configuration, Accounting, Performance, Security) defined by the international organization for standardization (ISO) model. This interface supports the communication of network service management related to the network functions.
The O2-interface plays a role in the communication between O-cloud and the SMO, helping the workload and resource allocation of the O-RAN system. Security of this channel is extremely important \cite{DBLP:journals/corr/abs-2201-06080}.
The Open Fronthaul Management Plane Interface helps the Network Management Systems (NMSs) and the DUs with the RU administrations. The A1-interface is in charge of connecting the Near-RT RIC and Non-RT RIC to transmit the data  stored in the SMO (e.g. the training data for the ML model, the enrichment information of O-RAN, etc.)
The E2-interface is used to connect the near-RT RIC with the E2 nodes including CU, DU, and eNB to control and optimize resource allocation. Generally, E2-interface supports all 3GPP-defined layers and protocols. O-RAN alliance is working on the standardization of all the above-mentioned interfaces, of which the  need of security enhancement is needed.

The openness of the proposed revolutionary architecture 
brings not only the advantages but a vastly larger attack surface. With the concept of “secure by design,” the
discussion of security issues in ORAN architecture and the related technologies is
extremely important in the early development stages of ORAN.

The remainder of the article is organized as follows.
Section 2. defines our review scope in terms of the identified attack surfaces in O-RAN. Section 3. provides more detailed descriptions of possible attacks corresponding to the attack surfaces and further presents the existing mitigation methods. Section 4. gives the literature review. Finally, in section 5., the conclusions and recommendations are presented.

\section{Problem definition} 

This project would analyze the possible attack surfaces of the current O-RAN framework. From the O-RAN architecture itself to the technologies including virtualization and ML that are intended to be covered. The potential threats and recommended mitigation methods are also provided.

We identify four risk areas of O-RAN according to a specific major aspect of O-RAN framework based on the taxonomy proposed by Dudu Mimran et al \cite{DBLP:journals/corr/abs-2201-06080}. The following are the four risk areas we would look into in-depth:

\subsection{Architectural openness}

Openness is the core concept of O-RAN. The open framework offers several advantages such as the reduction of operational costs and the opportunities for small and medium-sized enterprises, which further accelerate the development of RAN technologies. Nonetheless, openness comes along with more interfaces and larger variances of security levels, which opens new attack surfaces in the O-RAN network.

\subsection{Cloud and Virtualization}

In the O-RAN network, the RAN functions are disaggregated and deployed on the edge clouds utilizing related cloud and virtualization technologies. Multiple processes with different functionalities are allowed to be run on a single unit with virtual machines (VMs) or containers. These procedures could be monitored and managed in a centralized way, allowing a more flexible resource allocation and the reduction of operational costs. Nonetheless, the explosion of attack methods targeting the vulnerabilities in the cloud, including virtualization and containerization, are highly transferable to the virtualized RAN architecture as well. Therefore, this would also be a key part of our research on which we will focus, including Image Manipulation Attacks, Guest-to-Guest Attacks, Guest-to-Hypervisor Attacks, and Inconsistent Secure Policies as illustrated in the fig. 2.
\begin{figure}[h]
  \centering
  \includegraphics[width=\linewidth]{ {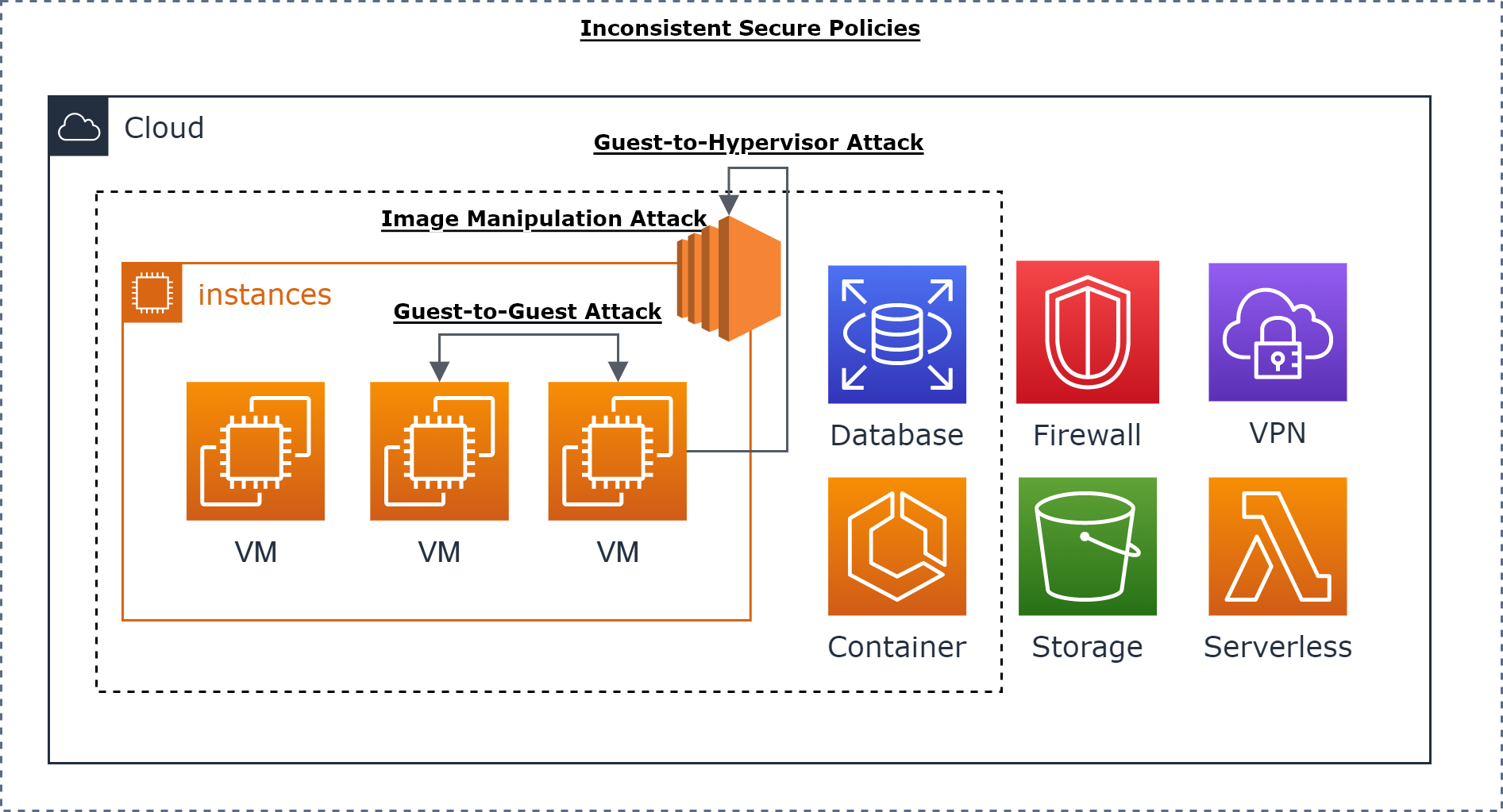} }
  \caption{Cloud and Virtualization Technology Attack Surfaces.}
  \Description{A diagram that pinpoints the four attack surfaces of cloud and virtualization that our research will focus on.}
\end{figure}

\subsection{Network Slicing}
The implementation of network slicing also plays an important role in the O-RAN context based on the cloud-native nature of O-RAN. Network slicing divides the physical networks into several virtual network slices, allowing services for specific Quality of Service (QoS) requirements and offering better resource allocations and isolation for end-to-end logical networks. Although bringing the above advantages, network slicing technology also gives rise to security and privacy issues that should be addressed.

\subsection{Machine Learning}
Intelligence is another key objective in the O-RAN system in order to support complex cellular network management challenges such as resource scheduling, modulation classification, spectrum sensing, etc. 
O-RAN framework defines system-level intelligence units including SMO and RICs to manage the distributed and disaggregated components of the system to meet the Enhanced Mobile Broadband (eMBB), Ultra-reliable and Low Latency Communications (URLLC), and Massive Machine Type Communication (mMTC) requirements of 5G. These units enable the deployment of machine models and even third-party machine learning applications.\\

The main difference between our project and previous works is that we pay more attention to the areas of Cloud and Virtualization, Network Slicing, and Machine Learning. Furthermore, we study the existing attack methods and corresponding solutions or recommendations, 


\section{Results: Threats and Mitigations}

\subsection{Architectural openness}
O-RAN comprises several third Generation Partnership Project (3GPP) compliant logical functional units. Every unit could communicate with each other through open and standardized interfaces. The threats from the increasing numbers of stakeholders and vendors should be carefully considered as the attack points could be found along the whole supply chain and even to the third-party API providers. O-RAN is also open to attacks that target traditional RANs or other cellular protocols. The newly introduced channel might make O-RAN even more susceptible to the attacks such as the \textit{Passive eavesdropping} that extract sensitive data from the communication between UE and RU, the \textit{Radio jamming} which aims to disrupt the transmission of radio signals, and the \textit{Side-channel} attacks that steal the user information from the user based on the physical property. Proper guidelines, policies, and protocols must be defined and followed to secure the O-RAN. Last but not least, it is extremely important to adopt thorough and comprehensive security assessments (e.g. authentication)in the transmitting protocols of the interfaces to guarantee secure data transmissions between the units.

\subsection{Cloud and Virtualization}

As cloud computing is a broad field that is essentially an abstraction of all IT services. It is destined to become more and more versatile and includes more and more different features, it is technically impossible to analyze all aspects of it in our work. Therefore, we would like to focus on cloud computing services, which are likely to be the first major area of adoption for O-RAN architecture in the future. In this section, for every attack surface we mentioned, we would explain the concept, give an example of a related attack, provide mitigations, and further explain its implications for O-RAN.   

\subsubsection{Image Manipulation}\ \\
Applications are stored and executed from binary image files in virtualized environments. Some threats target the vulnerabilities in binary image files during image creation and execution. For example, attackers may provide malicious images containing older versions of OpenSSL and Unix Bash Shell to victims. Utilizing Shellshock (CVE-2014-6271) and HeartBleed (CVE-2014-0160) to gain illegal access.

Another vulnerability related to image creation is known as \textit{Template Images Cloning} \cite{grobauer2010understanding}. To provide nearly identical server images, template images are often used to save computational resources. However, a template image may have been manipulated so as to provide back-door access for an attacker. Cloning can also leak private data to other users. In Amazon EC2, users can provide template images for other users by turning a running image into a template. Without strict management and configuration, it may contain data that the user doesn't wish to make public.

For image-related vulnerabilities, we would like to suggest users update images to the latest versions automatically and use image filters and virtual scanners frequently. N{\"u}wa\cite{zhou2010always} is a tool that enables efficient and scalable offline patching of dormant VM images, while Mirage\cite{wei2009managing} is an image management system to remove confidential information and detect malicious images.

\subsubsection{Guest-to-Guest Attack}\ \\
In virtualized environments, applications \textbf{share the same resources} and may share storage, a CPU, or a host OS. Several threats target those resources as a means to attack other guest applications, which seem to be even more prevalent in the context of virtualization. Attack methods include communicating with other not mutually trusted guest applications, injecting malicious execution code, leaking information through side-channel attacks, denial of service (DoS) attacks on other guest applications, etc. 

One kind of Guest-to-Guest attack we would like to further dive into is known as the \textit{Flush+Reload side-channel attack} as illustrated in \cite{184415}. Due to a weakness in Intel x86 processors, the page sharing feature that allows for copy-on-write leads to unexpected information leakages. This attack focuses on the Last-Level-Cache (LLC) of the CPU which is shared among multiple cores on the same processor die. With shared pages, the attacker can ensure what is in and what is not in the cache hierarchy. Therefore, the attacker can monitor the victim's access to the memory line. The basic concept can be summarized as the following.

\begin{enumerate}[1.]
    \item Attacker flushes the shared LLC with the x86 instruction \texttt{clflush}
    \item Victim \textbf{[accesses / does not access]} memory.
    \item Attacker reloads with \textbf{[short / long]} reload time.
    \item Attacker infers which memory line is accessed by the victim and thus what instructions are executed
    \item Repeat and grind to extract more information
\end{enumerate}

Possible mitigations against the attack include restricting the usage of the \texttt{clflush} instruction and using newer CPUs, which feature non-inclusive memory caches. For the former case, it is claimed that use cases of the instruction are limited to the system function of maintaining cache coherence and improving program performance. Therefore, it is simply sufficient to limit \texttt{clflush} to the memory pages to which the process has write access. For the latter case, non-inclusive cache memories, where cache memories in L1 do not necessarily have to be in L2 in L3. Therefore, the attack becomes useless as evicting a block from LLC does not mean evicting it from the cache completely.     

Though this kind of attack may be harder to carry out on private clouds like O-RAN where it is a lot harder to get intimate knowledge of the other tenants running on the same infrastructure, it serves as a great example for us to see how related attacks can extract information (private keys, credentials, etc. ) from other O-RAN services and how we could implement defenses against them.

\subsubsection{Guest-to-Hypervisor Attack}\ \\
Similar to Guest-to-Guest attacks, the Guest-to-Hypervisor attack also targets shared resources between the hypervisor and guests. However, since the hypervisor also acts as a single point of failure, these kinds of attacks focus more on compromising the hypervisor in order to have more power to carry out further attacks. Existing threats especially aim at harming the integrity of the hypervisor, DoS attack on the hypervisor, or trying to avoid the monitoring of VMMs (Virtual Machine Monitor), etc.

To more clearly explain the concept above, we would like to show a case in which a Guest-to-Hypervisor Attack is carried out in order to pull off a Guest-to-Guest Attack. In this scenario, the attacker tries to stall the live migrations \cite{206172} of VMs on the same hypervisor so that it has time to carry out the Flush+Reload Attack. As mentioned above, in order to carry out a Flush+Reload Attack, the attacker needs to have sufficient knowledge of the neighboring tenants in order to carry out the attack successfully. Since the attack usually takes approximately a few ten minutes ($\approx$ 26 minutes) to complete, performing live migrations periodically can greatly reduce side-channel attacks \cite{206172}. Based on this knowledge, the guest-to-hypervisor attack tries to create dirty pages and cause bus contention deliberately in order to slow down the process of live migrations, which increases the length of the attack time span and, thus, the success rate. 

Possible mitigations for the attack include setting resource usage limits on the process of live migrations and allocating more resources for the migrating VM to make it harder for the attacker VM to gather enough resources in order to stall the migrating VM for a long enough duration. 

As ORAN infrastructure gradually adopts a cloud-native approach, more and more different functions will be virtualized. Therefore, how the hypervisor manages all of these complex relations and defends against malicious attackers would be an important area of concern when designing ORAN infrastructures.

\subsubsection{Inconsistent Secure Policies}\ \\
Since the core value of ORAN is its openness, misconfiguration resulting from human errors where humans can be the developer of components, integrators, engineers, and operators represent the highest risks in the system. Moreover, as the complexity of cloud-based assets grows, the range of attacks on these assets and the stakes concerned has also been on the rise. This includes the exploitation of misconfigurations of cloud databases, storage, and computational units. Not to mention, the perimeterless nature of cloud-based systems also makes them vulnerable to malicious actors. 

Many of these issues mentioned above already exist in our public cloud service providers and, therefore, should also be kept in mind when designing future ORAN architectures. One example is the widespread misconfiguration of the well-known cloud storage service AWS S3 \cite{10.1145/3274694.3274736}. Studies that try to target the specific service and scan for as many buckets as possible have revealed a daunting statistic. According to the work, out of all the 240,000 S3 buckets found, 14\% of them are publicly-listable, 11\% of them are readable, and 2\% of them are writable. Readable buckets are likely to leak private information such as user data or private credentials, while writable buckets can be even more dangerous because of possible defacement, web resource injections, and domain name trust takeovers, which could prove dangerous for website owners and users simultaneously.

Possible mitigations for the attack include all kinds of education, awareness, verification tools, and others. To give some examples, for bucket owners, a tool to safely check the access policy of their buckets (including sensitive text and file extension analysis) would be extremely helpful in ensuring the deployment of safe configurations. On the other hand, for website users,  we could also provide a tool that checks for the existence of writable S3 buckets and refuse from loading those dangerous web resources. Though solving the human problem regarding security by and large is far from possible, we would like to emphasize the importance of it in the process of designing open systems, including O-RAN. 

\subsection{Network Slicing}
 By and large, the concept and technologies used in network slicing are similar to those utilized in network virtualization, and so are the attacks and mitigation methods. However, it can be seen as an independent technology that deserves special attention. We would like to discuss the network-slicing- specialized attack surfaces based on the survey by Ruxandra F.Olimid et al. \cite{olimid20205g}

\subsubsection{Slice life cycle}\ 
\begin{description}[style=unboxed, leftmargin=0cm]
    \item [Preparation.] During this phase, a slice template defining the framework and the configurations of a slice is created. Attacks that are similar to the \textit{Image Manipulation} might take place. A poorly designed or tampered with slice template might affect all the slices built from it. As such, the secure cryptographical protocols must be adopted during slice design and the integrity of each slice template must be checked before deployment. The disclosure of the template content should be prevented.
    
    \item [Instantiation, configuration, and activation.] A new slice of resources and network functions would be built based on the template in this phase. One should keep his/her eye open to the threats from APIs. Adversaries might also forge a slice in this phase. To mitigate the possible attacks, one should only use secure APIs on their slice and adopts TLS for data transmission. For better defense, the APIs should also support monitoring their own traffic log files by legal means.
    
    \item [Run time.] The slice is in use and is vulnerable to the changes of the slice (e.g. mitigation of slice configuration, permission setting, resource allocation) in this phase. One should also be careful of data exposures and the availability damage of the slice. Due to frequent access of the slice and the broad and various users, there are a large variety of attacks during the preceding phase. Overall, to defend against these attacks, the importance of authenticity and integrity verification should be emphasized to deny the adversary instance. A more detailed discussion of attacks and mitigation methods during the slice operation is discussed in the \textit{Intra-slice security} and \textit{Inter-slice security} sections.
    
    \item [Decommissioning.] The slice would be eliminated in this phase when its mission is completed. Inaccurate deactivation of slices might lead to the exposure of sensitive data and malicious resource consumption. Destruction of the archive and the de-allocation of network resources of the decommissioned slice should be performed.
\end{description}

Other aspects of threats concern the interactions between different slices or between entities within the same slice (e.g., resources, operation, management).
\subsubsection{Intra-slice security}\ \\
This section discusses the threats that target and only influence a specific slice. These kinds of threats include unauthorized access to slice resources, attacks targeting the interconnection between slices, and the exploitation of slice resources. End-to-end security of all communications must be considered to defend against these kinds of attacks. Primary and secondary authentication of the tenants should be adequately designed.

\subsubsection{Inter-slice security}\ \\
This section discusses the threats that target and influence other slices running on the same physical resource. For example, a performance attack could occupy all the resources on one physical infrastructure and the attacker could execute other types of attack in the meantime due to a lack of services for important security processes. Unauthorized access to sensitive data from a low-security slice is also possible. The guarantee of slice isolation, especially isolation from the slice with sensitive data, should be guaranteed to prevent these attacks because security is only as strong as the weakest link. Secure management is also recommended. 

\subsection{Machine Learning}

\begin{figure}
    \centering
    \includegraphics[width = \linewidth]{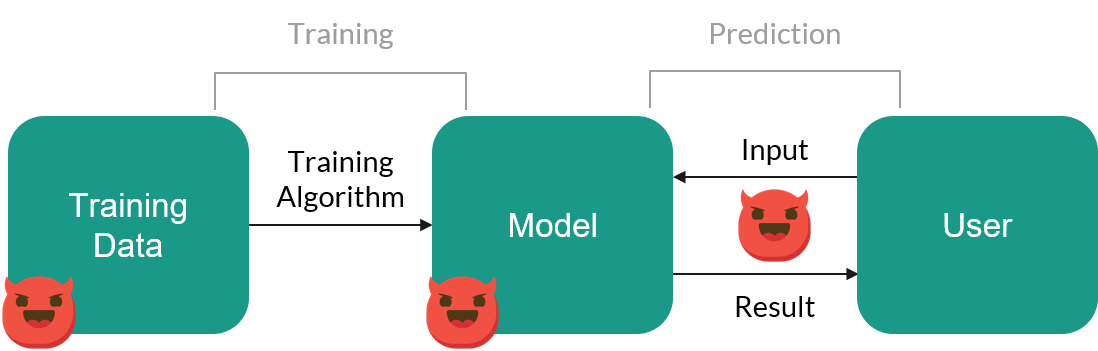}
    \caption{ML workflow and vulnerabilities}
    \label{fig:ml_overview}
\end{figure}

Figure \ref{fig:ml_overview} shows a simplified ML workflow. The basic idea of ML is to take in a large volume of data and their true labels in the training phase. The model would adjust its parameters to a state where the predicted results and the true labels have minimum prediction error. Once the model is well-trained, users can query the model with input data, the model will compute the corresponding predicted result. ML prediction has been used in a wide range of fields such as image classification, natural language processing, marketing, etc., and gained considerable success. In recent years, the usage of ML in wireless communication has started to gain researchers' attention. 5G architecture also incorporates ML to improve the performance of the network. However, ML is highly susceptible to even simple adversarial attacks. Without proper consideration, ML approaches may expose the network to new threats.

\subsubsection{ML Threats}\ \\
We follow the work of He, Yingzhe, et al. \cite{he2020towards} and respectively discuss the three aspects of ML threats.

\begin{description}[style=unboxed, leftmargin=0cm]
    \item [Training data.] As the concept of ML is to delve into the data and seek intelligent information. The feasibility and performance of an ML model are highly dependent on the training data set. \textit{Poisoning attack} exploits this weakness by maliciously inserting mislabelled data into the training data set, degrading the quality of training data and further affecting the quality of the model. This attack could make the ML model unusable, violating the availability and integrity of the model. Since the training data is crucial for model training, the collection and management of data are also valuable assets. \textit{Model inversion attack} allows attackers to restore information in the training data from the prediction results. Information such as the data memberships (whether the given data is in the training data set) and data properties could be stolen with this attack, leaking trade secrets and directly causing a loss to the operating company.
    
    \item [Trained model.] The training process takes a lot of time and costs in data collecting, storage, and computing, the well-trained model is highly valuable to a company. \textit{Model extraction attack} allows attackers to recover the structure, parameters, decision boundaries, etc., of the ML models by querying the model and duplicating an approximate model without the need to perform additional training. The duplication of ML models will cause enormous damage to the model training company assets or can be useful for further attack.
    
    \item [Inputs and results of predictions.] In the prediction phase, the input data can also be exploited by the attackers to affect the outcome of the prediction. \textit{Evasion attacks} carefully modified the input data to produce adversarial examples, the change is unperceived by humans while the model will wrongly classify the adversarial example. 
\end{description}

The threats in the ML systems above expose O-RAN to new weaknesses. These attacks not only harm the operating company's interest but also make the wireless network unusable. Performing poisoning attacks on the intelligent components in O-RAN could lower the efficiency in network managing, or lead to loss of availability. The model inversion attack may also leak users' privacy. The evasion attack can not only be performed to affect a victim's prediction result, a misbehaving UE can also intentionally input adversarial examples to fool the ML models whose functionality relies on UEs' input. In the traffic steering use case, the attacker can fool the ML model, making its own connection handover frequently between multiple cells \cite{https://doi.org/10.48550/arxiv.2201.06093}. This could lead to the exhaustion of resources because the overhead of handover is expensive.

\subsubsection{mitigations}\ \\
There are several techniques to make the ML model more tolerant of malicious inputs. \textit{Adversarial training} includes the adversarial examples in the training data set, making the ML model gain inherent resistance against adversarial inputs. \textit{Defensive distillation} uses the knowledge extracted from the ML model to improve its own resistance against adversarial examples \cite{https://doi.org/10.48550/arxiv.1511.04508}. \textit{Concept drift} continuously monitors the performance of ML model to detect adversarial examples while \textit{adversarial detection} focuses on the input of the ML model.

Collecting users' data for model training may also be a privacy issue, cryptographic methods such as homomorphic encryption, multiparty computation, and zero-knowledge argument schemes can be applied to protect privacy while still allowing the computation for learning. \textit{Differential privacy} protects the privacy by adding noises in the prediction result to protect training data.

\section{Related work}
DongHyun Je et al. \cite{je2021toward} analyzed the potential new threats due to the
introduction of new technologies in wireless communication systems from an
extensive aspect of 6G. Other groups focus on the ORAN concept and provide
more specific research. Dudu Mimran et al. \cite{DBLP:journals/corr/abs-2201-06080} proposed an ontology and developed a process for ORAN security evaluation. A threat analysis of ORAN
based on the proposed process is conducted. However, their research lacks a
survey of possible solutions corresponding to the identified threats. CT Shen et al.
\cite{shen2022security} studied several pieces of research about information securities in 5G and
ORAN yet they did not present a detailed analysis of the reviewed cases from a
higher point of view.

\section{Conclusion and future work}

In this project, four main attack surfaces of O-RAN, including (1) Architectural openness, (2) Cloud and Virtualization, (3) Network slicing, and (4) Machine Learning are identified and followed by further study of the possible solutions to the corresponding threats. Several examples of existing attacks and mitigation methods are also discussed. Our results could serve as a baseline and a guideline for operators or other implementers in building cyber defense strategies for O-RAN networks. Although we agree that a vast attack surface is created along with the introduction of the openness concept in O-RAN systems. We believe that, through careful implementation of the concept of security by design in every fragment and framework of O-RAN since its early-development stage, O-RAN would evolve into a robust and secure system in the long term. The enhancement of the security level of O-RAN still needs the contribution of the community. More thorough and more practical research of attacks and certified defense methods should be conducted in the future. Continuous updates and re-evaluation according to the corresponding specifications updated of O-RAN are also needed.

\bibliographystyle{ACM-Reference-Format}
\bibliography{main}

\end{document}